\documentclass[aps,prb,twocolumn,superscriptaddress,nofootinbib]{revtex4-1}
\usepackage{graphicx}
\usepackage{dcolumn}
\usepackage{bm}
\usepackage{amsmath}
\usepackage{amssymb}
\usepackage{latexsym}
\usepackage{epsfig}
\usepackage{amsbsy}
\usepackage{array}
\usepackage{amssymb}
\usepackage{setspace}
\usepackage{bm}
\usepackage{braket}
\usepackage{prettyref}
\usepackage{comment}
\usepackage{mathrsfs}
\usepackage{color,xcolor}
\usepackage{comment}
\usepackage{blkarray}
\usepackage{subfigure}
\usepackage{float}
\usepackage{tikz}
\usetikzlibrary{positioning}
\usetikzlibrary{shapes.geometric}
\usetikzlibrary{shapes.misc}

\def\sint{\ifmmode{- \!\!\!\!\!\! \int}
    \else{\hbox{$- \!\!\!\! \int \ $}}\fi}

\allowdisplaybreaks

\begin{document}
\title{Topological energy gaps in the [111]-oriented InAs/GaSb and GaSb/InAs core-shell nanowires}
\author{Ning Luo}
\affiliation{Key Laboratory for the Physics and Chemistry of Nanodevices and Department of Electronics, Peking University, Beijing 100871, China}
\author{Gaohua Liao}
\affiliation{Key Laboratory for the Physics and Chemistry of Nanodevices and Department of Electronics, Peking University, Beijing 100871, China}
\author{Lin-Hui Ye}
\affiliation{Key Laboratory for the Physics and Chemistry of Nanodevices and Department of Electronics, Peking University, Beijing 100871, China}
\author{H. Q. Xu}
\email{hqxu@pku.edu.cn; hongqi.xu@ftf.lth.se}
\affiliation{Key Laboratory for the Physics and Chemistry of Nanodevices and Department of Electronics, Peking University, Beijing 100871, China}
\affiliation{Division of Solid State Physics, Lund University, Box 118, S-221 00 Lund, Sweden}
\date{\today}

\begin{abstract}
The [111]-oriented InAs/GaSb and GaSb/InAs core-shell nanowires have been studied by the $8\times 8$ Luttinger-Kohn $\vec{k}\cdot\vec{p}$ Hamiltonian to search for non-vanishing fundamental gaps between inverted electron and hole bands. We focus on the variations of the topologically nontrivial fundamental gap, the hybridization gap, and the effective gap with the core radius and shell thickness of the nanowires. The evolutions of all the energy gaps with the structural parameters are shown to be dominantly governed by quantum size effects. With a fixed core radius, a topologically nontrivial fundamental gap exists only at intermediate shell thicknesses. The maximum gap is $\sim 4.4$ meV for GaSb/InAs and $\sim 3.5$ meV for InAs/GaSb core-shell nanowires, and for the GaSb/InAs core-shell nanowires the gap persists over a wider range of geometrical parameters. The intrinsic reason for these differences between the two types of nanowires is that in the shell the electron-like states of InAs is more delocalized than the hole-like state of GaSb, while in the core the hole-like state of GaSb is more delocalized than the electron-like state of InAs, and both features favor stronger electron-hole hybridization. Since similar features of the electron- and hole-like states have been found in nanowires of other materials, it could serve as a common rule to put the hole-like state in the core while the electron-like state in the shell of a core-shell nanowire to achieve better topological properties.
\end{abstract}


\maketitle

\section{Introduction} 


Topological insulators\cite{kane2005z,moore2007topological,fu2007topological} are the class of materials which are insulating in the bulk while possess time reversal symmetry protected metallic states on the surface.  Until now, all identified topological insulators are either 3D bulk materials\cite{fu2007topological,zhang2009topological,xia2009observation} or 2D quantum well structures\cite{bernevig2006quantum,konig2007quantum,roth2009nonlocal,knez2011evidence,du2015robust,knez2011evidence,du2015robust}. Although it is not clear whether 1D topological insulator truly exists, the InAs/GaSb and GaSb/InAs core-shell nanowires\cite{ganjipour2012carrier,ganjipour2015,rieger2015misfit} are considered as possible candidates based on the identification of the topological properties in the corresponding quantum well structures\cite{knez2011evidence,du2015robust}. Here, the key feature is that the conduction band bottom of InAs is lower than the valence band top of GaSb, thus offering the possibility of band inversion, the characteristics of most topological insulators. Further, for the nanowire to be insulating inside, a net fundamental gap must exist over the whole momentum space. In principle, such a gap can form through the anti-crossing of the electron and hole bands, and, if the ordering of these bands is inverted, then the fundamental gap is topologically nontrivial. 

Experimentally, the observation of the helical edge state in the InAs/GaSb quantum well\cite{knez2011evidence,du2015robust} strongly suggests that a topologically nontrivial fundamental gap could also exist in the corresponding nanowire structures. On the other hand, ambipolar transport has been realized in the GaSb/InAs core-shell nanowires\cite{ganjipour2012carrier} starting from a threshold thickness of about 5 nm of the InAs shell, indicates that the fundamental gap is near to closure in these particular samples. Later, a quantum dot system has been realized with a GaSb/InAs core-shell nanowire and the effect of the interaction between electron quantum dot states and hole quantum dot states has been observed,\cite{ganjipour2015} indicating a clear occurrence of an energy overlap of InAs conduction band states and GaSb valence band states in the core-shell nanowire. 

\begin{figure}[b]
\centering
\includegraphics[width=8.5cm]{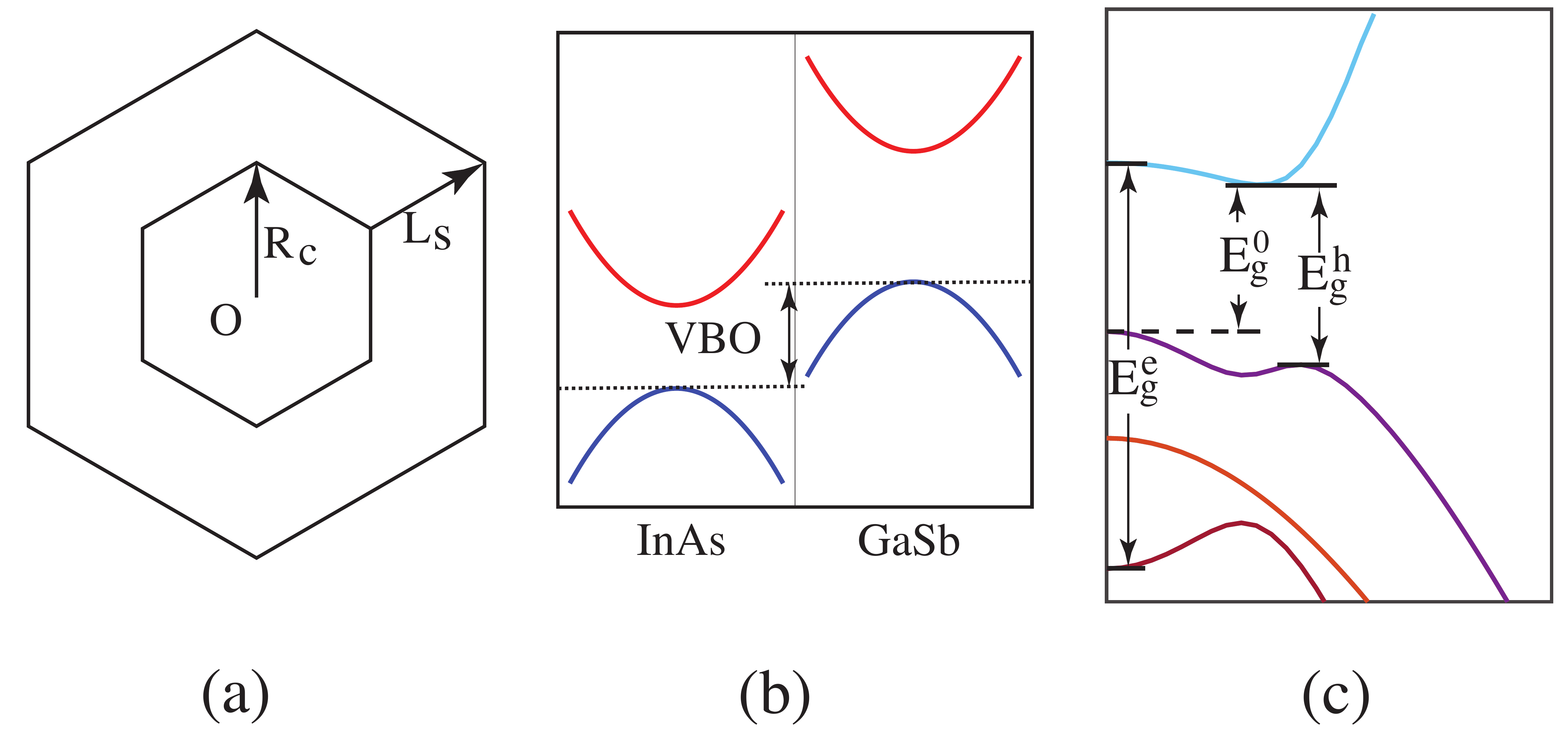}
\caption{(a) (Color online) Schematic cross-sectional structure of the core-shell nanowires. $R_c$ is defined as the radius of the core and $L_s$ the thickness of the shell. (b) Band alignment of bulk InAs and GaSb. The valence band offset (VBO) is defined as the energy difference betweent the valence band maxima of the two materials. (c) Illustration of the topologically nontrivial fundamental gap ($E_g^0$), the hybridization gap ($E_g^h$) and the effective gap ($E_g^e$) in a band structure of a core-shell nanowire with band inversion.\label{Fig:Theory1}}
\end{figure}

On the theory side, as illustrated in Fig.~\ref{Fig:Theory1}, three different types of energy gaps can, in principle, exist in an InAs/GaSb or a GaSb/InAs core-shell nanowire with band inversion: (i) the topologically nontrivial fundamental gap $E_g^0$ mentioned above, (ii) the hybridization gap $E_g^h$ resulting from the anti-crossing of the inverted electron and hole bands which, by definition, does not include gap between normally ordered states,  and (iii) the effective gap $E_g^e$, i.e., the energy difference between the lowest electron-like band and the highest hole-like band at the $\Gamma$ point. In presence of band inversion, the effective gap becomes negative. Here, of particular importance is the evolution of the hybridization gap with the structural parameters. Since in the Brillouin zone all hybridization gaps are formed only around individual points, they are often covered up by continuous band states elsewhere. To form a topologically nontrivial fundamental gap, therefore, at least one of the hybridization gaps must remain open over the entire Brillouin zone. 

The electronic structure of the [001]-oriented InAs/GaSb and GaSb/InAs nanowires have been calculated by Kishore \textit{et al.}\cite{2012PeetersInAs/GaSb} These authors have explored the variation of the band structure as a function of the structural parameters, such as the radius of the core and the thickness of the shell. Although it is not specifically targeted on the topological properties, some useful information can already be extracted from the work. However, a few important questions  still remain open. First, the results reported so far are only for nanowires with the [001] orientation. It is not clear how they differ for nanowires oriented along other crystallographic directions, especially along the experimentally most relevant [111] direction. Second, although a useful effective-gap map has been offered, of direct relevance to the topological properties is the topologically nontrivial fundamental gap for which no such a map has ever been presented. Third, despite that both the InAs/GaSb and GaSb/InAs core-shell nanowires are experimentally accessible, most theoretical results are for the former type while the latter type has only been briefly discussed. Therefore, more detailed studies of the GaSb/InAs core-shell nanowires are needed.

It is the these important questions which have led to the work presented in this paper. In the paper, we use the $\vec{k}\cdot\vec{p}$ theory to calculate the band structures of the InAs/GaSb and GaSb/InAs core-shell nanowires oriented along the [111] crystallographic direction. We analyze band inversion and the evolutions of the three types of energy gaps with the structural parameters. In particular, we offer detailed maps of the topologically nontrivial fundamental gaps over a wide range of structural parameters for both the InAs/GaSb and GaSb/InAs core-shell nanowires. A fully quantitative comparison between the results obtained for InAs/GaSb and GaSb/InAs core-shell nanowires reveals that although these two types of nanowires are qualitatively similar, the latter type is more favorable concerning topological properties.

The rest of this paper is organized as the following: In Section \ref{sect:Theory} we briefly review the theoretical methodology of this work.   The main results for the InAs/GaSb and GaSb/InAs core-shell nanowires are presented in Sections \mbox{III A} and \mbox{III B}, respectively. In particular, we will illustrate in detail how the quantum size effects govern the formation of the topologically nontrivial fundamental gap. Section \mbox{III C} is denoted to the evolution of the effective gap with the change in the geometrical structures. The sign of the effective gap serves as an indicator to the topological property of an existing fundamental gap. In Section \mbox{III D}, we present a quantitative comparison between the topological properties of the InAs/GaSb and GaSb/InAs core-shell nanowires, and illustrate that the latter structure favors stronger electron-hole hybridization and is thus more preferred when concerning topological properties. Finally, we conclude this work in Section IV.

\section{Methodology\label{sect:Theory}}

Various methods, including density-functional-theory (DFT)\cite{ning2015remote,ning2013first,cahangirov2009first,srivastava2013first,dos2010diameter}, tight-binding (TB) theory\cite{persson2004electronic,persson2006electronic,persson2006electronic2,niquet2007effects,niquet2006electronic,luisier2006atomistic,liang2007performance,paul2010performance,persson2008electronic,liao2015electronic2,liao2015electronic1}, and the $\vec{k}\cdot\vec{p}$ theory\cite{he2009performance,kishore2014electronic,2012PeetersInAs/GaSb,2006Lassen111InPInAs,kohn1955theory,luttinger1956quantum,citrin1989valence,xia1991effective,2012PeetersInAs/GaSb,2012Peeters6v8,lassen2004exact,2006Lassen111InPInAs,Peeters2008optical,kishore2010electronic}, have been used to calculate the band structure of nanowires. Of the three, DFT is the only first-principle method which is free from any adjustable parameters, and therefore is often the best choice for electronic structure calculations. Specific to the energy gap properties of the nanowires concerned in this work, however, DFT is incompetent because it suffers from the well-known  ``band gap problem'', in that one major component of the fundamental gap, the derivative discontinuity of the exchange-correlation functional, is missing in the popular local density approximation (LDA) and generalized gradient approximations (GGA). Consequently, the DFT band gaps are typically 0.1 eV to a few eV too small, with the range of the errors being already on the orders of magnitude larger than the topologically nontrivial fundamental gaps interested in this work. Besides, DFT is relatively costly. Typically, its use is limited to a few hundred atoms which is often not sufficient to treat nanowires of realistic sizes.

Both TB and the $\vec{k}\cdot\vec{p}$ theory can avoid the above DFT problems at the expense of using empirical parameters. Comparatively,  $\vec{k}\cdot\vec{p}$ is computationally more efficient, and is therefore particularly suited for the exploration of nanowires with large sizes. As the energy gaps involve both the valence band top and the conduction band bottom, the $8\times 8$ $\vec{k}\cdot\vec{p}$ Hamiltonian needs to be employed. The Hamiltonian is formulated in the basis of the light hole (LH), heavy hole (HH), spin split-off (SO) band states, as well as the conduction band electron (EL) state. Then, by including  up and down spins the size of the basis set is doubled. In this work, most band and Luttinger parameters are taken from Vurgaftman \textit{et al.}\cite{vurgaftman2001band}. The only exception is the Kane energy $E_p$ of GaSb which has been changed to 24.76 eV as suggested  by Foreman\cite{foreman1997elimination} to avoid spurious solutions. Throughout this paper, the valence band offset (VBO, see Fig.\ref{Fig:Theory1}) is fixed to 0.56 eV\cite{claessen1986pressure,ekpunobi2005curvature,sai1978optical} and the energy zero is set at the top of the valence band of bulk InAs.

\begin{figure*}[th]
\includegraphics[width=17cm]{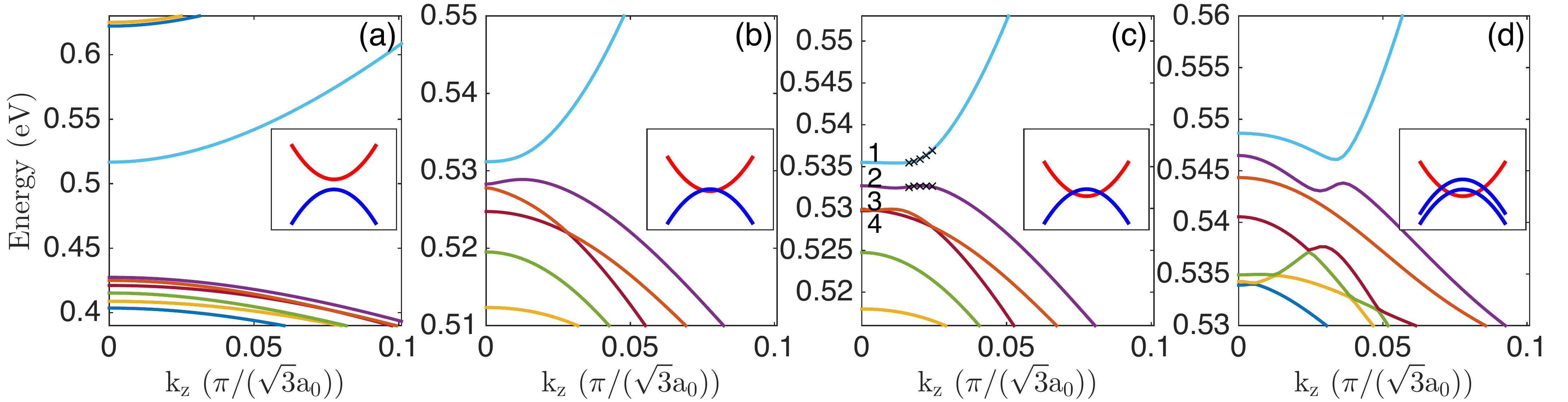}
\caption{(Color online) Band structures of the InAs/GaSb core-shell nanowires with a fixed core radius of $R_c =9.0$ nm. From left to right, the shell thicknesses $L_s$ are (a) 2.0 nm, (b) 5.3 nm, (c) 5.9 nm, and (d) 9.0 nm.  In (c),  the lowest conduction band state and the three highest valence band states  at the $\Gamma$-point are labeled by integer numbers. For these bands, their character analyses are presented in Fig.~\ref{Fig:InAsGaSb_3_Spinor}. The crosses placed on the lowest conduction band and the highest valence band in (c) indicate the states for which the wave functions are plotted in Fig.~\ref{Fig:InAsGaSb_GaSbInAs_Kzwave}. The inset in each panel indicates schematically the relative positions of the conduction and valence bands in the calculated band structure shown in the panel. \label{Fig:InAsGaSb_Band}}
\end{figure*}

Another slight difference from the earlier work is that we use hexagonal boundary for all nanowires which is more relevant to the experiments than a circular boundary used in the earlier work. For the purpose of study of the core-shell nanowires oriented along the [111] crystallographic direction in this work, the principal axis of the Hamiltonian has to be rotated from the [001] to the [111] crystallographic direction. Then the Hamiltonian matrix is constructed by the finite element method. Finally, through diagonalization, band energies and band state wave functions are obtained. All technical details can be found elsewhere.\cite{Luo2016}

\section{Results and Discussion}

In this section, we focus on the evolutions of the topologically nontrivial fundamental gaps in the InAs/GaSb and GaSb/InAs core-shell nanowires with the structural parameters. In a core-shell nanowire, the total band structure is essentially formed by the two sets of bands from InAs and GaSb, subject to their mutual interactions and the effects of quantum confinement. Especially, the conduction bands of InAs and the valence bands of GaSb are close in energy, and it is the hybridization between them which may offer the fundamental gap we are interested in. On the other hand, the valence bands of InAs and the conduction bands of GaSb are either too low or too high in energy and are essentially irrelevant to the formation of the gap. Throughout this paper, bands mostly of the InAs characteristics are termed ``electron-like'' due to their positive parabola-like dispersions. Similarly, bands mostly of GaSb characteristics are termed ``hole-like'' due to their negative parabola-like dispersions. 

\subsection{Topologically nontrivial fundamental gaps in the InAs/GaSb core-shell nanowires\label{sect:InAs}}

\subsubsection{Band structure analysis}

By fixing the InAs core radius while varying the GaSb shell thickness, we have calculated the band structures of four representative nanowires. Fig~\ref{Fig:InAsGaSb_Band} shows the calculated band structures of the InAs/GaSb core-shell nanowires with a fixed InAs core radius of $R_c=9.0$ nm.  The GaSb shell thicknesses considered in Figs.~\ref{Fig:InAsGaSb_Band}(a) to \ref{Fig:InAsGaSb_Band}(d) are $L_s=2.0$ nm, 5.3 nm, 5.9 nm, 9.0 nm, respectively. Since the most important features are located near the $\Gamma$-point, the bands are drawn only for about 1/10 of the half Brillouin zone.
Fig~\ref{Fig:InAsGaSb_Band}(a)  shows the results of the calculations for the nanowire with the smallest shell thickness of $L_s=2.0$ nm. Here, a large energy gap of 0.09 eV is found at the $\Gamma$-point. Band character analysis (not shown) reveals that the bottom conduction band is electron-like and the top valence band hole-like. Therefore, band ordering is normal and the effective gap is positive. The 0.09 eV energy gap, in this case also the fundamental gap, is topologically trivial.  Besides, both the lowest conduction band and the highest valence band have smooth parabolic shape, indicating small hybridization due to the large energy separation. 

\begin{figure}[tb]
\includegraphics[width=8.5cm]{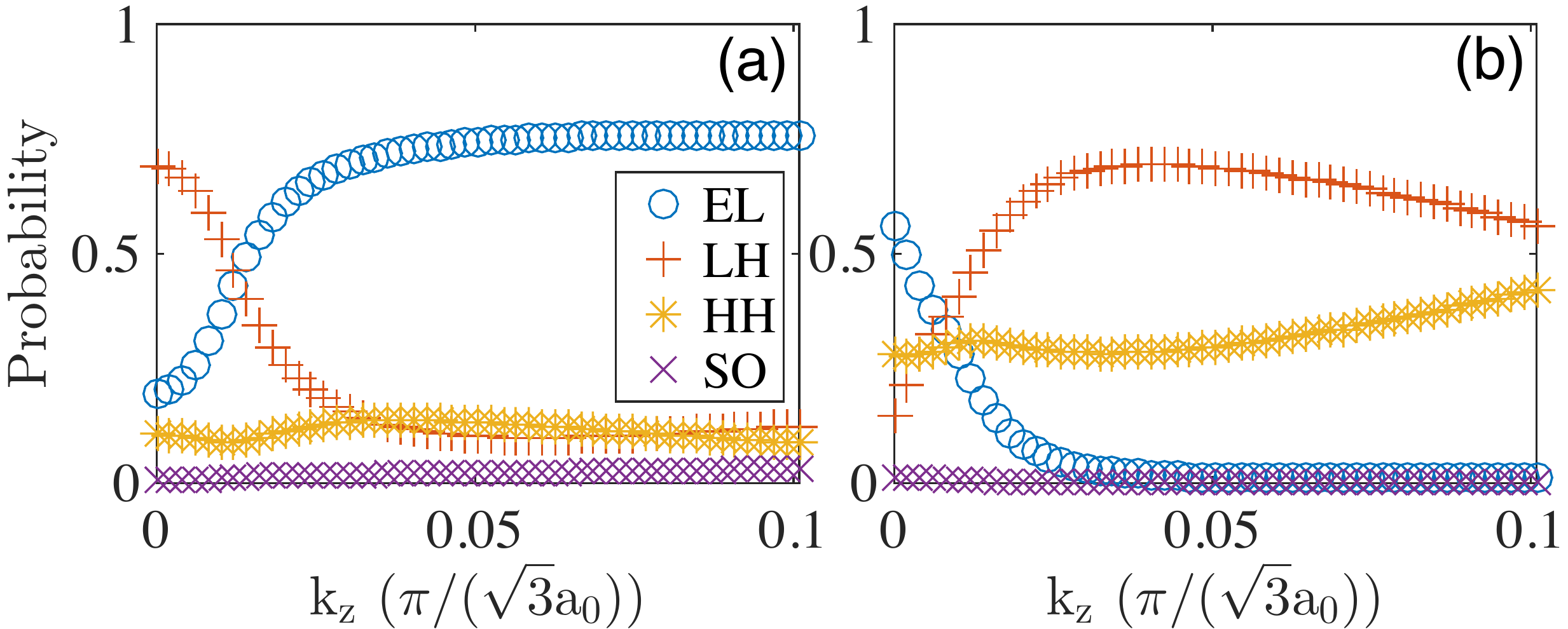}
\caption{(Color online) Band characters of (a) the lowest conduction band and (b) the highest valence band shown in Fig.~\ref{Fig:InAsGaSb_Band}(b) for the InAs/GaAs core-shell nanowire with the core radius $R_c=9.0$ nm and the shell thickness $L_s=5.3$ nm.\label{Fig:InAsGaSb_12_Spinor}}
\end{figure}

Figure~\ref{Fig:InAsGaSb_Band}(b) shows the results of the calculations for the nanowire with the shell thickness of $L_s=5.3$ nm. Here it is seen that there is still an energy gap, but the size is only 2.3 meV. Note that the different energy scales are used in Fig.~\ref{Fig:InAsGaSb_Band}(b) and Fig.~\ref{Fig:InAsGaSb_Band}(a). The conduction band bottom and the valence band top, referred as the critical points, are not at the $\Gamma$-point, but shift slightly away from it. For Fig.~\ref{Fig:InAsGaSb_Band}(b), we have picked the  lowest conduction band and the highest valence band for which the band characters are shown in Fig.~\ref{Fig:InAsGaSb_12_Spinor}. From the $\Gamma$ to the critical points, the  conduction band is hole-like while the valence band electron-like. Therefore,  band ordering is inverted which leads to a negative effective gap. The 2.3 meV gap is thus an authentic hybridization gap which offers a net, topologically nontrivial fundamental gap at this shell thickness. 
According to Fig.~\ref{Fig:InAsGaSb_12_Spinor}, band inversion is only limited to the small region around the $\Gamma$-point. On the other hand, beyond the critical points the electron character gradually resumes in the lowest conduction band while the hole character gradually resumes in the highest valence band, so that band ordering returns to normal. The smallness of the band inversion region in Fig.~\ref{Fig:InAsGaSb_12_Spinor} implies that $L_s=5.3$ nm is close to the threshold thickness of the shell for band inversion to happen.

\begin{figure}[tb]
\includegraphics[width=8.5cm]{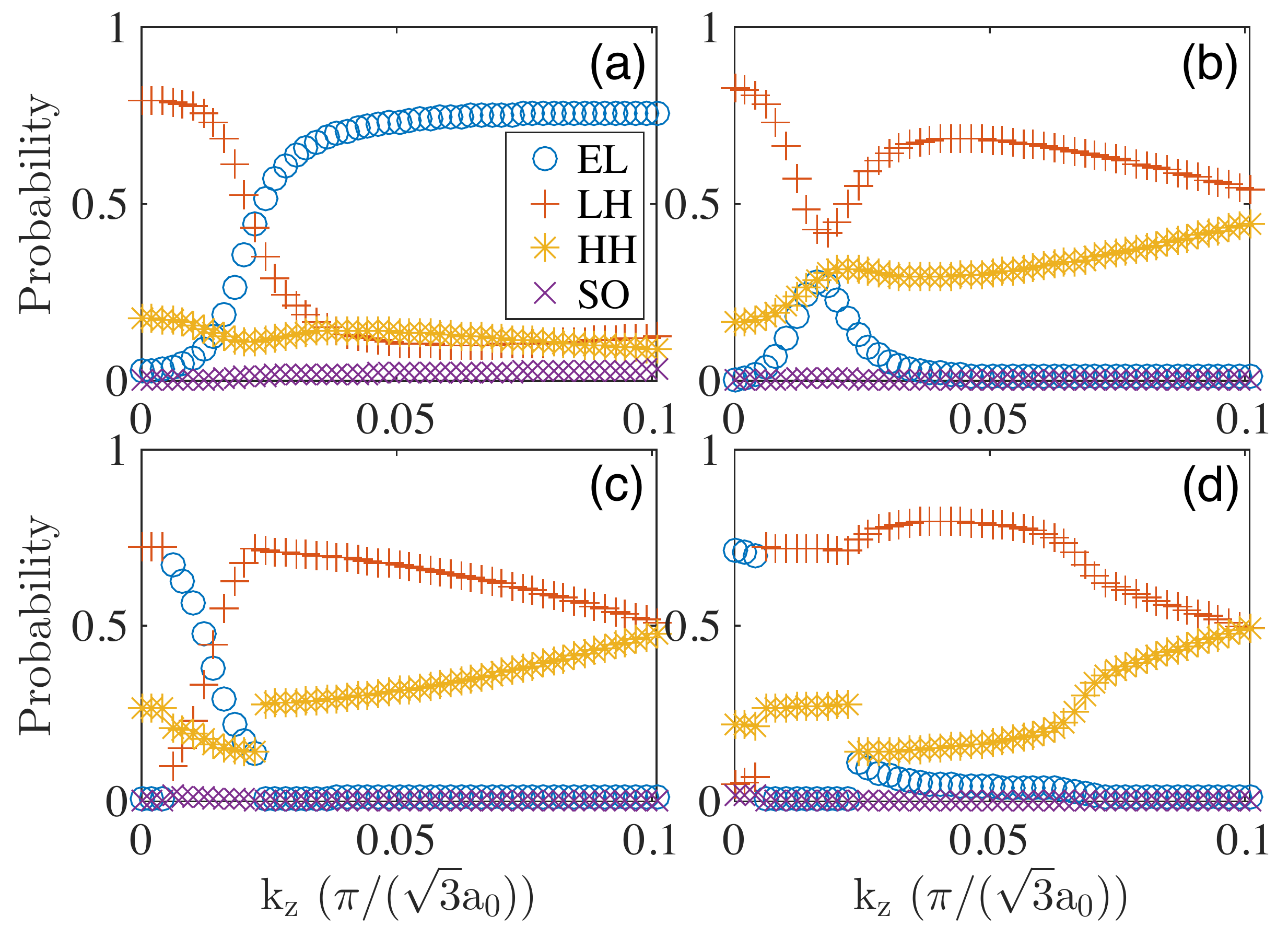}
\caption{(Color online) Band characters of (a) band 1, (b) band 2, (c) band 3, and (d) band 4 shown in Fig.~\ref{Fig:InAsGaSb_Band}(c) for the InAs/GaAs core-shell nanowire with the core radius $R_c=9.0$ nm and the shell thickness $L_s=5.9$ nm. \label{Fig:InAsGaSb_3_Spinor}}
\end{figure}

Figure~\ref{Fig:InAsGaSb_Band}(c) shows the results of the calculations for the nanowire with the shell thickness of $L_s=5.9$ nm. Here it is seen that there is still an energy gap of 2.7 meV, and the critical points shift further away from the $\Gamma$-point. It is also seen in Fig.~\ref{Fig:InAsGaSb_Band}(c) that the electron-like band penetrates into the hole band region and overlaps with the three highest valence bands. These four bands are labeled by numbers for which we show their band characters in Figs.~\ref{Fig:InAsGaSb_3_Spinor}. 
Let us first look at the band characters of the lowest conduction band as shown in Fig.~\ref{Fig:InAsGaSb_3_Spinor}(a). It is seen that the band characters are very similar to that shown in Fig.~\ref{Fig:InAsGaSb_12_Spinor}(a), implying that the two lowest conduction bands shown in Fig.~\ref{Fig:InAsGaSb_Band}(b) and Fig.~\ref{Fig:InAsGaSb_Band}(c) have similar topological properties. Namely, for both bands, the hole character dominates the region between the $\Gamma$-point and the critical points, beyond which the electron character resumes. 

Of the three labeled valence bands in Fig.~\ref{Fig:InAsGaSb_Band}(c), band 4 has the lowest energy which only slightly touches the tail of the electron-like band at the $\Gamma$-point. Correspondingly, in its band character analysis shown in Fig.~\ref{Fig:InAsGaSb_3_Spinor}(d),  the electron character shows up only in a tiny region around the $\Gamma$-point, while in the outside of the region, the hole character dominates.
For the remaining two valence bands, bands 2 and 3, their band characters shown in Fig.~\ref{Fig:InAsGaSb_3_Spinor}(b) and Fig.~\ref{Fig:InAsGaSb_3_Spinor}(c) both show substantial electron-hole mixing. The maximum  mixing is found around the ``crossing point'' where the electron-like band would cross the hole-like band as seen in Fig.~\ref{Fig:InAsGaSb_Band}(c). This is because hybridization is increased with decreasing energy separation and achieves the maximum near energy degeneracy.

We now reach at the analysis of the band properties of the nanowire with the largest shell thickness of $L_s=9.0$ nm as shown in Fig.~\ref{Fig:InAsGaSb_Band}(d). Here, an energy gap of 2.3 meV is observable at $k_z\sim 0.04$ $\pi/\sqrt{3}a_0$.  Band character analysis (again not shown here) reveals that band inversion is still limited to the vicinity of the $\Gamma$-point. Therefore, the effective gap is negative and the 2.3 meV energy gap is an authentic hybridization gap.  Nevertheless, at this shell thickness the electron-like band falls so deeply in the valence band region  that the hybridization gap is covered up by the three high-lying valence bands. Consequently,  there is no net fundamental gap in Fig.~\ref{Fig:InAsGaSb_Band}(d).

\subsubsection{Quantum size effects}

The evolutions of the energy gaps from Fig.~\ref{Fig:InAsGaSb_Band}(a) to Fig.~\ref{Fig:InAsGaSb_Band}(d) reflect the fundamental physics of quantum confinement in the core-shell nanowires. In a pure semiconductor, three major consequences are expected from quantum confinement: (i) the valence top is pushed down,  (ii) the conduction bottom is pushed up, and therefore (iii) the band gap is increased.
In a core-shell nanowire, such quantum size effects are operative in both the core and the shell. To simplify our analysis, let us neglect the interface interactions so that the two sets of bands from the core and the shell can be analyzed separately. That is to say, if the core radius is fixed, we basically fix the set of bands from the core and only analyze the variation of the bands from the shell, and vice versa. Specific to Fig.~\ref{Fig:InAsGaSb_Band},  bands from the InAs core are fixed, and we only consider the variation of the bands from the GaSb shell with different levels of the quantum size effects. 

As shown in Fig.~\ref{Fig:Theory1}(b), the energy difference between the valence band tops of bulk GaSb and InAs, the so-called valence band offset (VBO),  is 0.56 eV. Counting the 0.42 eV band gap of InAs, the valence band top of GaSb is therefore 0.14 eV higher than the conduction band bottom of InAs.  Since the hole band is above the electron band, band ordering is intrinsically inverted.
For the InAs/GaSb core-shell nanowires, when the InAs core is fixed at $R_c=9.0$ nm and the GaSb shell is set to $L_s=2.0$ nm, the quantum size effects  in the shell are much stronger than in the core. At this small shell thickness, the GaSb valence band top has been pushed down so much that it falls below the InAs conduction band bottom, i.e., the band ordering becomes normal. Therefore, the 0.09 eV fundamental gap in Fig.~\ref{Fig:InAsGaSb_Band}(a) is topologically trivial. 

As the shell thickness is increased, the quantum size effects are gradually released, which causes the GaSb valence bands to move up. At $L_s=5.3$ nm, the valence band top of GaSb has just passed the conduction band bottom of InAs so that the intrinsic band inversion of the two bulk materials resumes. The small hybridization gap  in Fig.~\ref{Fig:InAsGaSb_Band}(b) comes from the anti-crossing of the inverted bands. Besides, from the inset to Fig.~\ref{Fig:InAsGaSb_Band}(b), it is  seen that these two bands  only slightly overlap which explains why the critical points are close to the $\Gamma$-point. 
When $L_s$ increase to 5.9 nm, the GaSb bands move further up. From the inset to Fig.~\ref{Fig:InAsGaSb_Band}(c), the band inversion region expands, pushing the critical points further away from the $\Gamma$-point. Finally, when $L_s$ increase to 9.0 nm, the GaSb valence bands have moved so high in energy, that several valence bands overlap with the electron-like conduction band as illustrated in the inset to in Fig.~\ref{Fig:InAsGaSb_Band}(d). As a result, the hybridization gap is completely covered up by the valence bands.

\begin{figure}[tb]
\includegraphics[width=8.5cm]{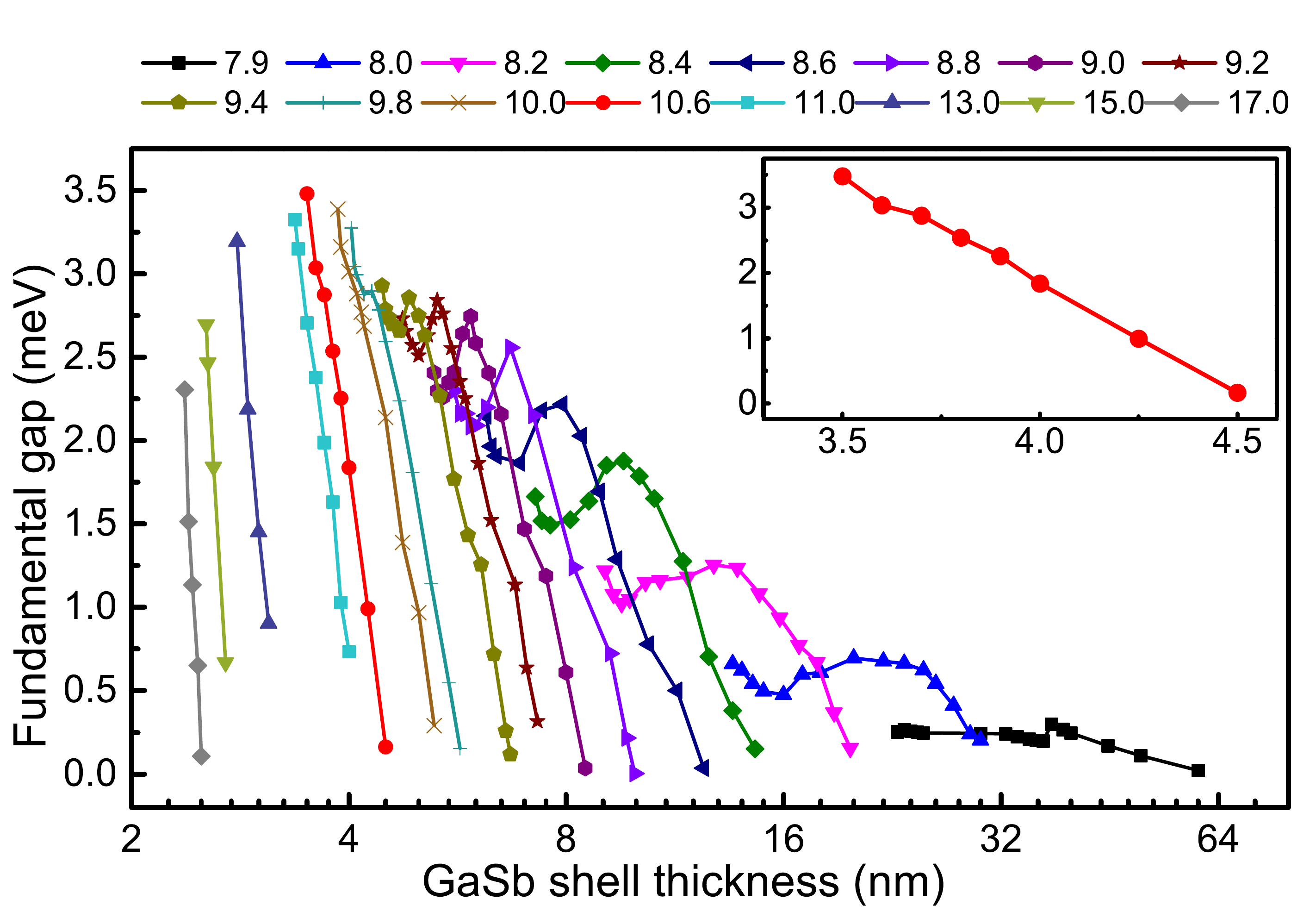}
\caption{(Color online) Map of the topologically nontrivial fundamental gap of the InAs/GaSb core-shell nanowire extracted from the band structure calculations over a large range of structural parameters. Each curve in the figure presents the variation of the topologically nontrivial fundamental gap with increasing shell thickness at a fixed core radius. The curve in the inset highlights the result for $R_c =10.6$ nm at which the maximum gap is found in the InAs/GaSb core-shell nanowire.
\label{Fig:InAsGaSb_minigap_map}}
\end{figure}

From the above analysis, we see that a net, topologically nontrivial fundamental gap exists only at intermediate shell thickness. If the shell is too thin, then the effective gap is positive and the fundamental gap is topologically trivial. On the other hand, if the shell is too thick, then although the effective gap turns negative, the electron-like band overlaps with too many valence bands, which causes the hybridization gap to be covered up. Thus, no fundamental band gap would exist.

Figure~\ref{Fig:InAsGaSb_minigap_map} shows the variations of the topologically nontrivial fundamental gap as a function of the shell thickness at different core radius (a map of the topologically nontrivial fundamental gap) of the InAs/GaSb core-shell nanowires extracted from the band structure calculations over a wide range of structural parameters. By closely examining the map, we find that the maximum topologically nontrivial fundamental gap present in the InAs/GaSb core-shell nanowire is 3.5 meV and occurs at $R_c=10.6$ nm and $L_s=3.5$ nm. For the core radius at this value of $R_c$, the evolution of the gap with increasing $L_s$ is shown in the inset to Fig.~\ref{Fig:InAsGaSb_minigap_map}.

\subsection{Topologically nontrivial fundamental gaps in the GaSb/InAs core-shell nanowires\label{sect:GaSb}}

In the above study of the band properties of the InAs/GaSb core-shell nanowires, we have simplified our analysis by neglecting the interface interactions and fixing the bands of the InAs core. In such a picture, it is the GaSb valence bands which {\it move up from below} and overlap more and more with the InAs conduction bands with increasing shell thickness. We now switch the core and shell materials, i.e., we put GaSb in the core and fix its geometrical structure, and put InAs in the shell and vary the shell thickness. Similarly, we assume that the GaSb bands are fixed and expect that the InAs conduction bands {\it move down from above} and overlap more and more with the GaSb bands with increasing shell thickness. 

\begin{figure*}[htb]
\includegraphics[width=17cm]{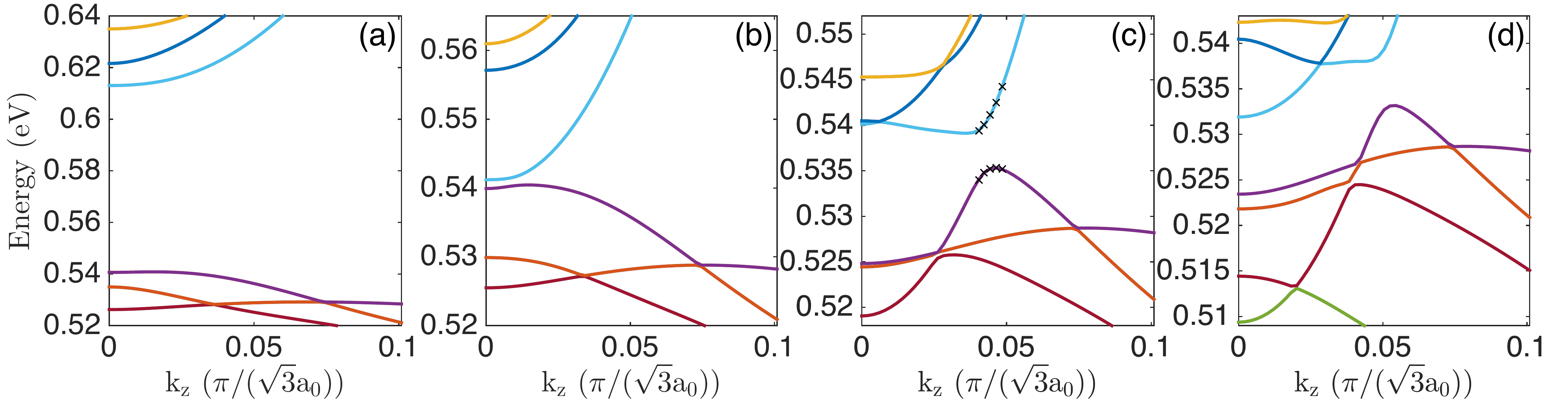}
\caption{(Color online) Band structures of the GaSb/InAs core-shell nanowires with a fixed core radius of $R_c=8.0$ nm. From left to right, the shell thicknesses $L_s$ are (a) 6.0 nm, (b) 8.87 nm, (c) 10.2 nm,  and (d) 11.0 nm.  The crosses placed on the lowest conduction band and the highest valence band in (c) indicate the states for which the wave functions are plotted in Fig.~\ref{Fig:InAsGaSb_GaSbInAs_Kzwave}. 
\label{Fig:GaSbInAs_Band}}
\end{figure*}

Figure~\ref{Fig:GaSbInAs_Band} shows the band structures of four representative GaSb/InAS core-shell nanowires with the same core radius of $R_c=8.0$ nm but different shell thicknesses. At the smallest shell thickness of $L_s=6.0$ nm, the conduction bands of InAs are so high in energy that the conduction band bottom is above the GaSb valence band top. Consequently, there is a finite energy gap as seen in Fig.~\ref{Fig:GaSbInAs_Band}(a) which is, however, topologically trivial. Although the results shown in Fig.~\ref{Fig:GaSbInAs_Band}(a) is similar to that in Fig.~\ref{Fig:InAsGaSb_Band}(a), there is a subtle difference that the highest valence band is not so smoothly parabolic but shows a weak ``camel back'' structure in the vicinity of the $\Gamma$-point, indicating that the hybridization of electron-like and hole-like bands in the GaSb/InAs core-shell nanowire  is stronger than in the corresponding InAs/GaSb core-shell nanowire, a point we will come back later.

From Fig.~\ref{Fig:GaSbInAs_Band}(b) to Fig.~\ref{Fig:GaSbInAs_Band}(d), it is seen that the increase of the shell thickness gradually releases the quantum size effect, causing the InAs conduction bands to gradually move down.  As shown in Fig.~\ref{Fig:GaSbInAs_Band}(b), at $L_s=8.87$ nm the electron-like band already falls below the highest valence band of the GaSb core and therefore band inversion starts to emerge. The anti-crossing of the inverted bands leads to a hybridization gap of 1.3 meV, with the critical points being very close to the $\Gamma$-point. Note that this hybridization gap is mostly covered up by the ``camel back'' structure, so that the net fundamental gap is only 0.5 meV. In this shell thickness, the fundamental gap is topologically nontrivial. 
Fig~\ref{Fig:GaSbInAs_Band}(c) shows that at $L_s=10.2$ nm, the electron-like band moves further down and its tail reaches the minimum energy of about 0.52 eV at the $\Gamma$-point. At the same time, the second lowest conduction band follows the same trend which now overlaps with the valence band at about 0.54 eV. The band inversion region is increased to almost half of the plotted $k_z$ vector region. The hybridization gap of 3.8 meV  comes dominantly from the anti-crossing of the lowest electron-like band with the highest valence band. 
Fig~\ref{Fig:GaSbInAs_Band}(d) shows that when $L_s$ further increases to 11.0 nm, the two lowest conduction bands have penetrated so deeply into the hole region that their tails reach about 0.51 eV and 0.53 eV, respectively. Due to the strong overlap between the electron-like and  hole-like bands, the hybridization gap from the first conduction band is fully covered up. At this shell thickness, there is no fundamental gap.

\begin{figure}[tb]
\includegraphics[width=8.5cm]{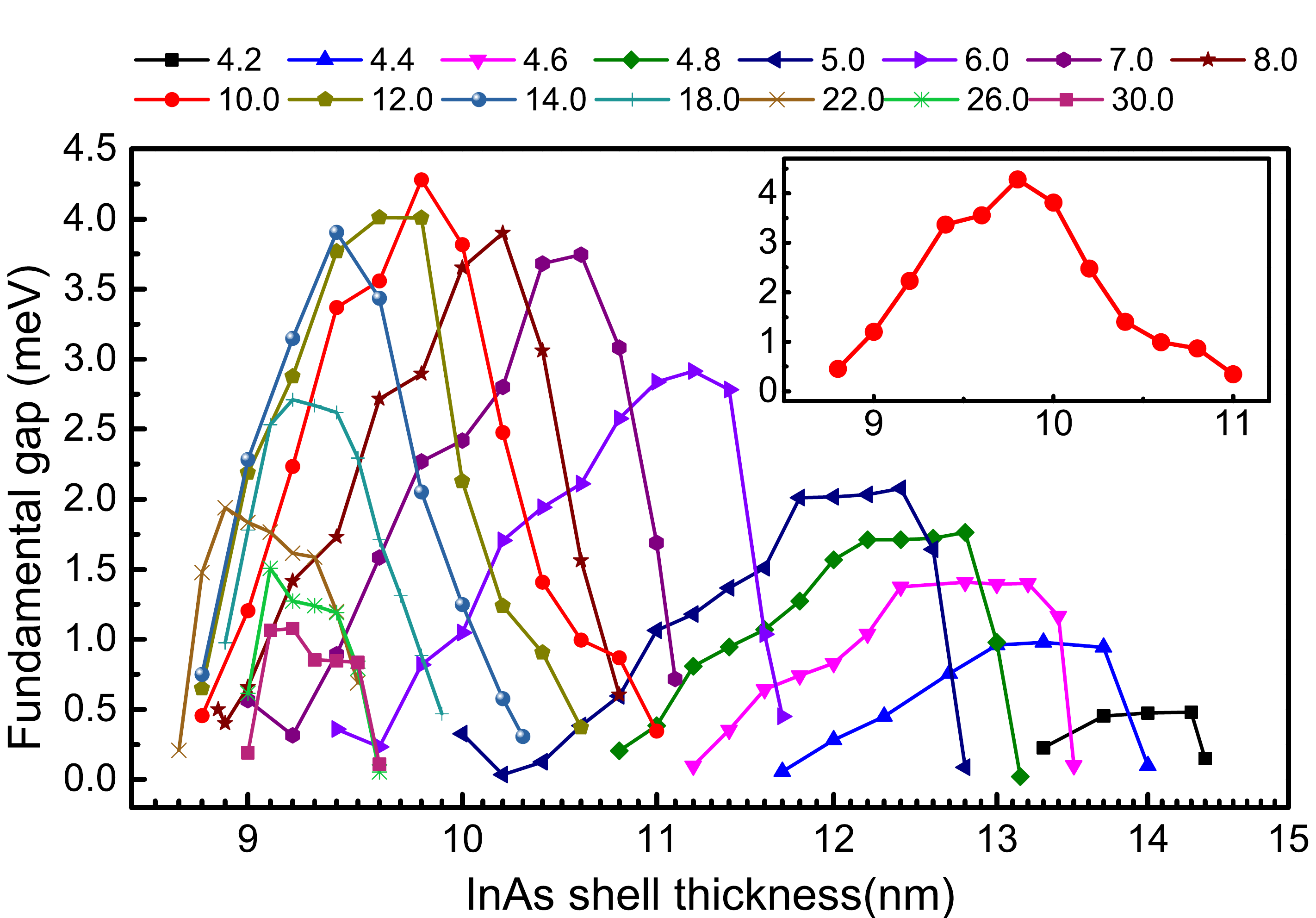}
\caption{(Color online) Map of the topologically nontrivial fundamental gap of the GaSb/InAs core-shell nanowire extracted from the band structure calculations over a large range of structural parameters. Each curve in the figure presents the variation of the topologically nontrivial fundamental gap with increasing shell thickness at a fixed core radius. The curve in the inset highlights the result for $R_c =10.0$ nm at which the maximum gap is found in the GaSb/InAs core-shell nanowire.\label{Fig:GaSbInAs_minigap_map}}
\end{figure}

Overall, the evolution of the energy gaps in the GaSb/InAs core-shell nanowires is similar to that of the InAs/GaSb core-shell nanowires. When the core radius is fixed to a sufficiently large values, a net, topologically nontrivial fundamental gap only exists at intermediate shell thickness. 

To search for the maximum topologically nontrivial fundamental gap in the GaSb/InAs core-shell nanowires, we have again calculated the band structures of the core-shell nanowires over a wide range of structural parameters and extracted a map for the topologically nontrivial fundamental gap from the calculations. The results are presented in Fig.~\ref{Fig:GaSbInAs_minigap_map}. It is seen that the maximum topologically nontrivial fundamental gap is 4.4 meV and occurs at the core radius of $R_c=10.0$ nm and the shell thickness of $L_s=9.8$ nm.  For the core radius at this value of $R_c$, the evolution of the gap with increasing shell thickness is highlighted in the inset to Fig.~\ref{Fig:GaSbInAs_minigap_map}. Here, it is seen that the maximum value of the topologically nontrivial fundamental gap occurs at an intermediate value of the shell thickness, as we discussed above.

\subsection{Effective gaps of the InAs/GaSb and the GaSb/InAs core-shell nanowires}

At presence of an finite fundamental gap, the sign of the effective gap of an InAs/GaSb or GaSb/InAs core-shell nanowire tells whether the gap is topologically trivial or nontrivial. Therefore, a systematic analysis of the effective gap offers another perspective of the topological properties of the nanowire. We now illustrate how the energies at the bottom of the electron-like bands and the top of the hole-like bands, both at $\Gamma$, vary with the structural parameters. To simplify our discussion,  we consider the case with a fixed  total radius of the core-shell nanowire, but varied sizes of the core and the shell. We will consider the representative results obtained from the calculations for the InAs/GaSb and GaSb/InAs core-shell nanowires with two total radii of 30 nm and 10 nm. 

\begin{figure}[tb]
\includegraphics[width=8.5cm]{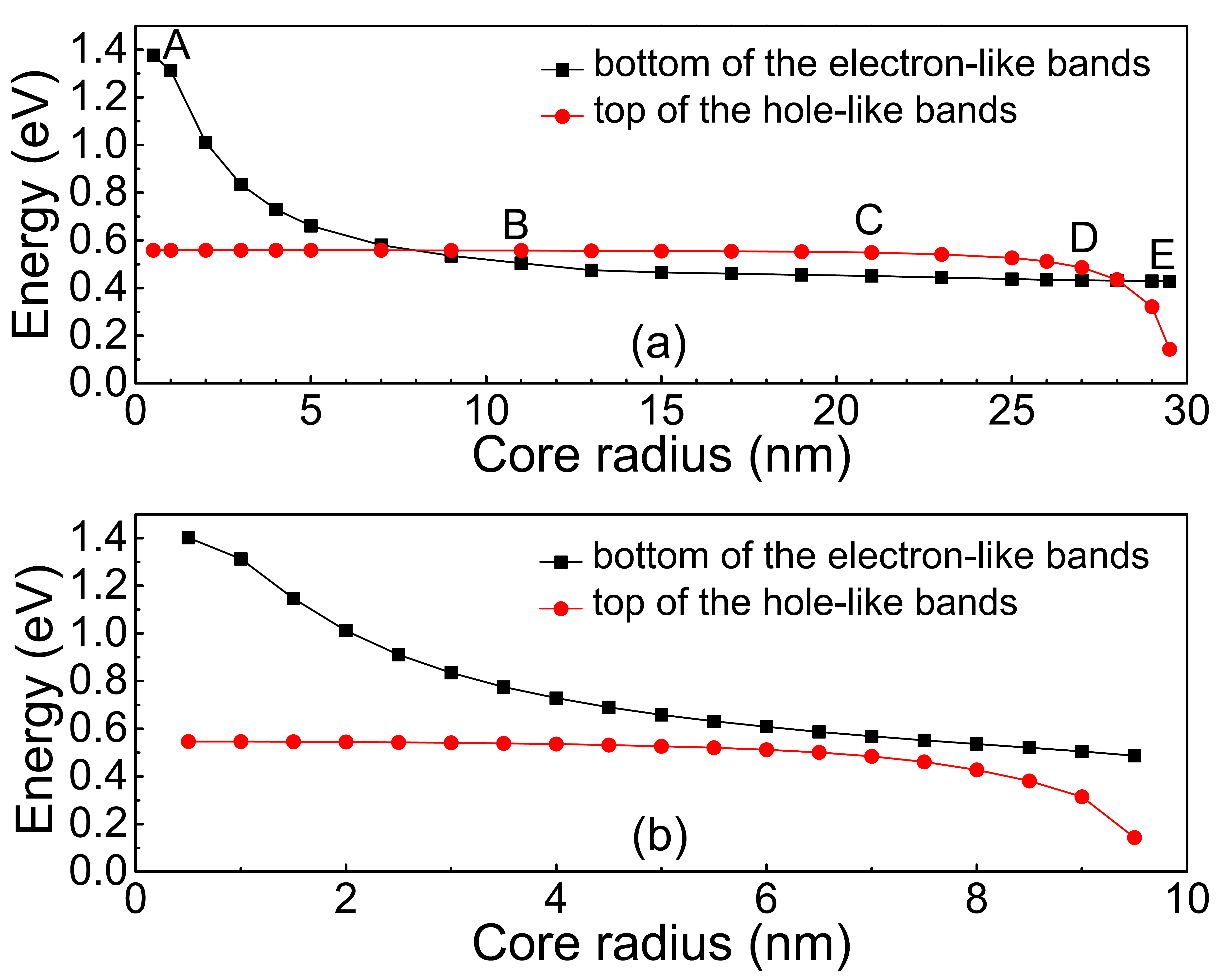}
\caption{(Color online) Energy positions at the bottom of the electron-like bands and the top of the hole-like bands of the InAs/GaSb core-shell nanowires. Here, the total sizes ($R_c+L_s$) of the nanowires are the same and are fixed at (a) 30 nm and (b) 10 nm, and the energy positions are plotted against the core radius $R_c$.\label{Fig:InAsGaSbR}}
\end{figure}

Figure~\ref{Fig:InAsGaSbR}(a) shows the band energies (at $\Gamma$-point) at the bottom of the electron-like bands and the top of the hole-like bands of the InAs/GaSb nanowire with a total radius of 30 nm. Now, we will choose five representative structural points for detailed analysis.  Point A in Fig.~\ref{Fig:InAsGaSbR}(a) corresponds to the all-shell limit where the core-shell nanowire becomes a freestanding pure GaSb nanowire. At this radius, the quantum size effects are so small that the effective gap is positive, and the size of the gap is close to the bulk value of GaSb, which is known to be 0.81 eV. On the other hand, point E in Fig.~\ref{Fig:InAsGaSbR}(a) corresponds to the  all-core limit where the core-shell nanowire becomes a freestanding pure InAs nanowire. Similarly, the effective gap is positive and is 0.42 eV, close to the bulk value of InAs.

In the structural parameter region between point A and point B in Fig.~\ref{Fig:InAsGaSbR}(a), the nanowire is mostly composed of the shell material, while the core is small. Therefore, the quantum size effects are mainly limited to the core, while the shell is not much affected. Correspondingly, in this region, the top of the hole-like bands derived dominantly from the GaSb shell barely changes, while the bottom of the electron-like bands  derived dominantly from the InAs core moves down quickly with increasing the core radius, due to the release of the quantum size effects. The effective gap is seen to changes from positive to negative when going from point A to point B. 
However, when going from point D and point E, the opposite is found. Here, the nanowire is mostly composed of the core material, while the shell is small. Therefore, the quantum size effects are mainly limited to the shell, while the core is not much affected. Correspondingly, the bottom of the electron-like bands does not vary much while the top of the hole-like bands moves down quickly. When the top of the hole-like bands passes the bottom of the electron-like bands, the effective gap changes sign from negative to positive.
In the region between point B and point D, relatively large absolute effective gaps are found, where the maximum value is found to appear at point C. In this region between point B and point D, all the effective gaps are negative ones and band inversion appear in the core-shell nanowires. At point C, which corresponds to $R_c=21$ nm and $L_s=9$ nm, the quantum size effects in both the core region and the shell region are small. Consequently, the bottom of the electron-like bands stays at energies close to the bulk value assigned to InAs and the top of the hole-like bands at energies close to the bulk value assigned to GaSb [see the schematic shown in Fig.~\ref{Fig:Theory1}(b)] and, thus, the obtained maximum absolute effective gap of 0.10 eV is not far from the intrinsic value of 0.14 eV as we would obtained for an ideal InAs/GaSb heterostructure as shown in Fig.~\ref{Fig:Theory1}(b).

We now explain that for band inversion to appear the total radius of the nanowire must be large enough. This is because, if the total radius of the nanowire is too small, there may not be a chance for the sizes of the core and shell to become sufficiently large at the same time. Fig~\ref{Fig:InAsGaSbR}(b) shows one such case where the total radius is only 10 nm. Here, the quantum size effects cannot be sufficiently small in both the core and the shell region. Therefore,  either the bottom of the electron-like bands is too high (if the InAs core is too small) or the top of the hole-like bands is too low (if the GaSb shell is too small), or both. In any of the cases, the effective gap is positive and no band inversion could appear.  This explains why in Fig.~\ref{Fig:InAsGaSbR}(b), there is a persistent fundamental gap which is always topologically trivial.

\begin{figure}[tb]
\includegraphics[width=8.5cm]{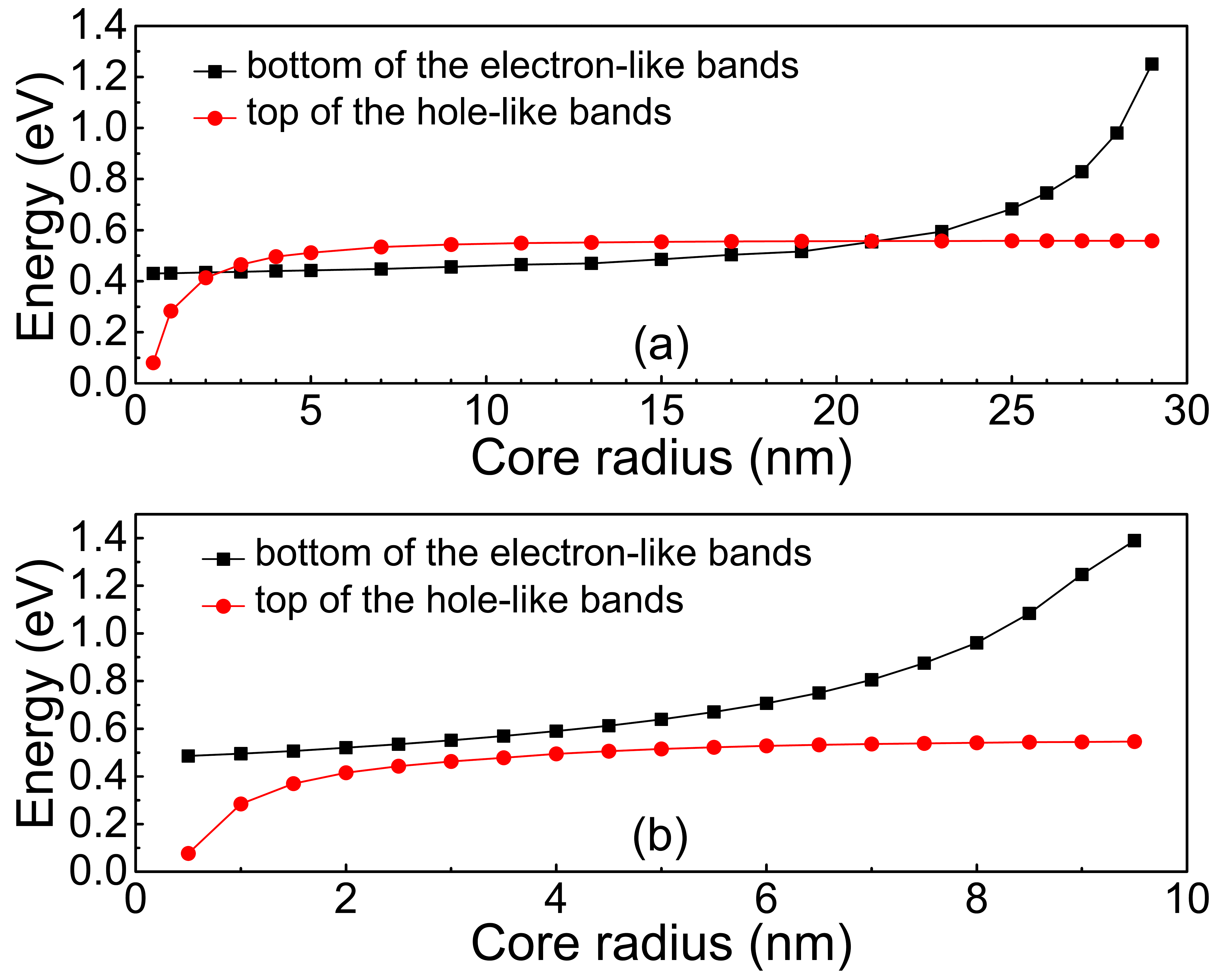}
\caption{(Color online) Energy positions at the bottom of the electron-like bands and the top of the hole-like bands of the GaSb/InAs core-shell nanowires. Here, the total sizes ($R_c+L_s$) of the nanowires are the same and are fixed at (a) 30 nm and (b) 10 nm, and the energy positions are plotted against the core radius $R_c$.
\label{Fig:GaSbInAsR}}
\end{figure}

Figure~\ref{Fig:GaSbInAsR} shows the corresponding calculations for the band energies (at $\Gamma$-point) at the bottom of the electron-like bands and the top of the hole-like bands of the GaSb/InAs core-shell nanowires. However, in the following, We shall bypass further analysis of the results, since Fig.~\ref{Fig:GaSbInAsR}(a) and Fig.~\ref{Fig:InAsGaSbR}(a), and Fig.\ref{Fig:GaSbInAsR}$(b)$ and Fig.~\ref{Fig:InAsGaSbR}$(b)$ are separately almost ``left-right symmetric''. Such symmetries indicate that the InAs/GaSb and GaSb/InAs nanowires are qualitatively similar, i.e.,  the effective gap of the core-shell nanowires mostly depends on the values of $R_c$ and $L_s$, but is not so sensitive to which of the two types of materials is in the core or in the shell. Next, we present a quantitative comparison between the topologically nontrivial fundamental gaps found in these two types of core-shell nanowires.

\subsection{Comparison between the InAs/GaSb and GaSb/InAs core-shell nanowires\label{sect:Comp}}

As far as topological properties are concerned, a nanowire structure is considered favorable if it possesses a large, topologically nontrivial fundamental gap, and this gap is robust against certain level of inaccuracy in experiments.  Applying the above criteria to the results shown in Fig.~\ref{Fig:InAsGaSb_minigap_map} and Fig.~\ref{Fig:GaSbInAs_minigap_map}, we find that the GaSb/InAs core-shell nanowires are more preferable than the InAs/GaSb core-shell nanowires because (i) the maximum gap is 4.4 meV for the GaSb/InAs core-shell nanowires which is larger than the maximum versus of 3.5 meV found for the InAs/GaSb core-shell nanowires and (ii) the topologically nontrivial gap generally persists in the GaSb/InAs core-shell nanowires over a wider range structural parameters. 

To better appreciate point (ii), let us first recall the results shown in Fig.~\ref{Fig:InAsGaSb_minigap_map} that for any given core radius, the gap only exists over a small range of shell thickness. For example, in the inset to Fig.~\ref{Fig:InAsGaSb_minigap_map}, corresponding to the core radius of $R_c=10.6$ nm, the gap is non-zero only for $L_s=3.5 \sim 4.5$ nm. 
Now let us consider the following more realistic question. Let us assume that the minimum acceptable gap to be 1 meV and require that the gap robustly exists over a $L_s$ range of at least 1 nm. Then, what is the range of the eligible core radius? It turns out that for a GaSb/InAs core-shell nanowire, any core radius within $R_c=4.6\sim 14.0$ nm allows such a gap to persists over the required shell range. On the other hand, for an InAs/GaSb core-shell nanowire, the corresponding core range is only $8.2\sim 10.2$ nm, which is much smaller. The central factor leading to the better topological properties of the GaSb/InAs core-shell nanowires is that hybridization in this type of nanowires is stronger, which we now elaborate below.

\begin{figure}[tb]
\includegraphics[width=8.5cm]{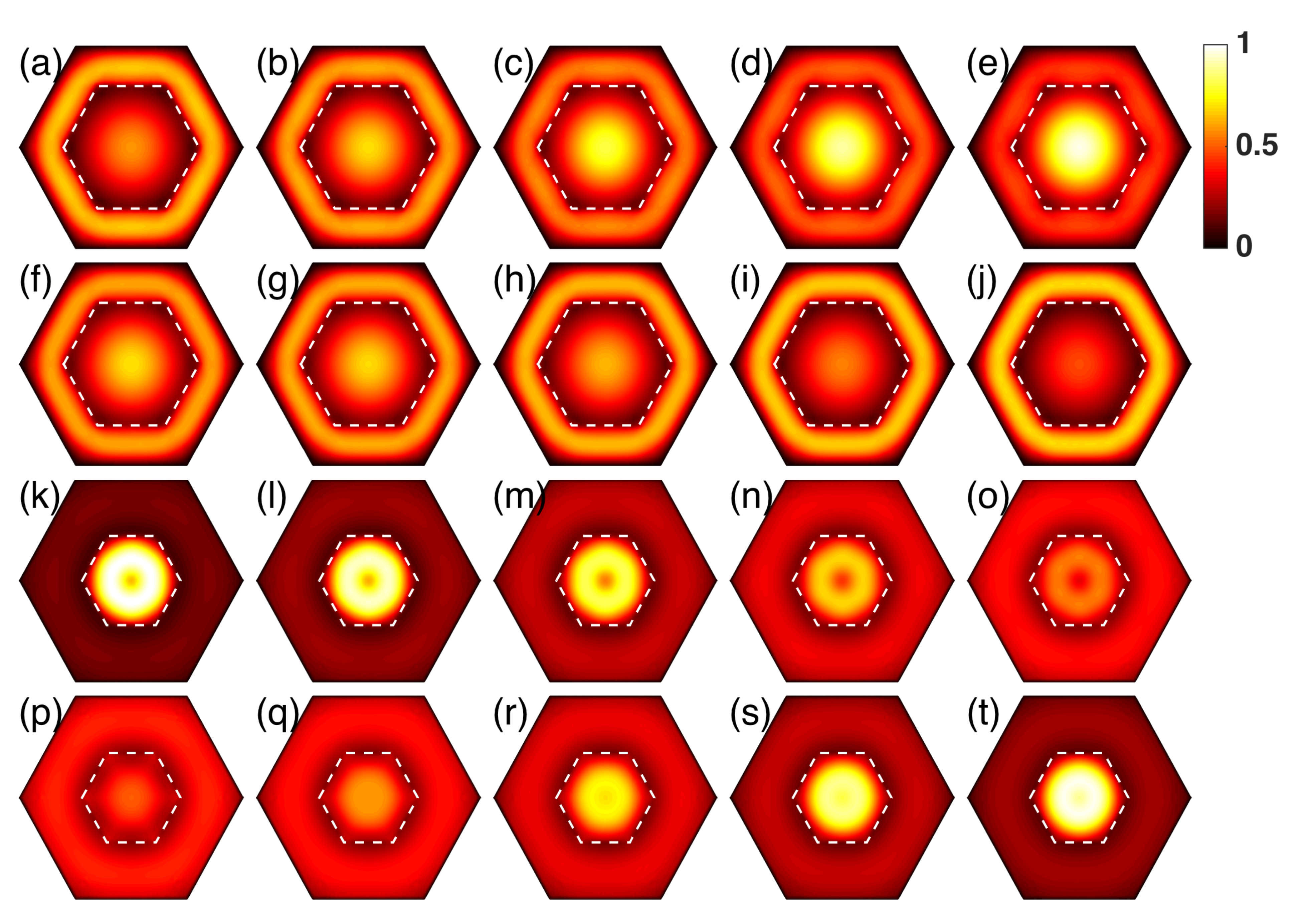}
\caption{(Color online) Probability distributions of the wave functions of the band states around the critical points marked by crosses in Fig.~\ref{Fig:InAsGaSb_Band}(c) and Fig.~\ref{Fig:GaSbInAs_Band}(c). (a)-(e) and (f)-(j) show the results for the five lowest conduction band states and the five highest valence band states of the InAs/GaSb core-shell nanowire with the core radius $R_c=9.0$ nm and the shell thickness $L_s=5.9$ nm, respectively. 
(k)-(o) and (p)-(t) show the results for the five lowest conduction band states and the five highest valence band states of the GaSb/InAs core-shell nanowire with the core radius $R_c=8.0$ nm and the shell thickness $L_s=10.2$ nm, respectively. Note that for comparison, the probability distribution values in all panels have been normalized against the highest value appeared in panel (k). The dashed lines in the figure mark the core-shell interfaces in the nanowires.\label{Fig:InAsGaSb_GaSbInAs_Kzwave}}
\end{figure}

In both Fig.~\ref{Fig:InAsGaSb_Band}(c) (For a InAs/GaSb core-shell nanowire) and Fig.~\ref{Fig:GaSbInAs_Band} (c) (For a GaSb/InAs core-shell nanowire), we have highlighted five $k_z$ points around each critical point of the hybridization gaps. For each of these $k_z$ points, we analyze the lowest conduction state and the highest valence state. The wave functions of these states have been plotted in Fig.~\ref{Fig:InAsGaSb_GaSbInAs_Kzwave}, in the following order of the lowest conduction band states of the InAs/GaSb core-shell nanowire (the first row), the highest valence band states of the InAs/GaSb core-shell nanowire (the second row), the lowest conduction band states of the GaSb/InAs core-shell nanowire (the third row), and the highest valence states of the GaSb/InAs core-shell nanowire (the fourth row).
In each row, the leftmost panel corresponds to the $k_z$ value closest to the $\Gamma$-point for which the band ordering is inverted. On the other hand, the rightmost panel corresponds to the $k_z$ value closest to the Brillouin zone boundary for which band ordering is normal. From left to right, we expect continuous change of the electron and hole characters.

Let us look at the first row which corresponds to the five bottom conduction states of the InAs/GaSb core-shell nanowire. Fig~\ref{Fig:InAsGaSb_GaSbInAs_Kzwave}(a) is dominated by a bright ring in the shell while the core is relatively dark, indicating the considered state contains a large contribution from the GaSb shell and small contribution from the InAs core. On the contrary, Fig.~\ref{Fig:InAsGaSb_GaSbInAs_Kzwave}(e)  is dominated by a bright disk covering the core region while the shell is dark, indicating the state contains a large contribution from the InAs core and a small contribution from the GaSb shell. Therefore, from Fig.~\ref{Fig:InAsGaSb_GaSbInAs_Kzwave}(a) to Fig.~\ref{Fig:InAsGaSb_GaSbInAs_Kzwave}(e), the band character changes from mostly hole-like to mostly electron-like.
The second row corresponds to the five top valence states of the InAs/GaSb core-shell nanowire. In Fig.~\ref{Fig:InAsGaSb_GaSbInAs_Kzwave}(f), both the core and the shell are bright, indicating strong electron-hole mixing. From Fig.~\ref{Fig:InAsGaSb_GaSbInAs_Kzwave}(f) to  Fig.~\ref{Fig:InAsGaSb_GaSbInAs_Kzwave}(j), however, the hole character gradually dominates the top valence band states.

The remaining two rows are for the GaSb/InAs core-shell nanowire. Essentially, we found the same trends as the InAs/GaSb core-shell nanowire, except that the core is dominated by the hole character while the shell is dominated by the the electron character. As expected, from Fig.~\ref{Fig:InAsGaSb_GaSbInAs_Kzwave}(k) to  Fig.~\ref{Fig:InAsGaSb_GaSbInAs_Kzwave}(o) the bottom conduction band state becomes more and more electron-like, while from Fig.~\ref{Fig:InAsGaSb_GaSbInAs_Kzwave}(p) to  Fig.~\ref{Fig:InAsGaSb_GaSbInAs_Kzwave}(t) the top valence band state becomes more and more hole-like.

The biggest differences between the two types of nanowires become clear when we separately look at the core and shell parts of the wave functions. Let us consider the shell first. If the shell is formed from GaSb, then the wave function in the shell is hole-like and shows a sharp circular ridge structure [see, e.g., Fig.~\ref{Fig:InAsGaSb_GaSbInAs_Kzwave}(a)], indicating that the hole-like state is very localized in the shell. Alternatively, if the shell is formed from InAs, then the wave function in the shell is electron-like and is much broader in space [see, e.g., Fig.\ref{Fig:InAsGaSb_GaSbInAs_Kzwave}(o)], implying that in the shell the InAs electron-like wave function is more delocalized then the GaSb hole-like wave function.
For the core similar analysis can be performed. If the core is formed from InAs, then the wave function in the core is electron-like and shows a strong peak around the center of the nanowire [see e.g. Fig.\ref{Fig:InAsGaSb_GaSbInAs_Kzwave}(e)]. On the other hand, if the core is formed from GaSb, then the wave function in the core is hole-like and does not show a peak but a dip around the center of the nanowire the wave function [see e.g. Fig.\ref{Fig:InAsGaSb_GaSbInAs_Kzwave}(k)], so that the maximum of the hole state is pushed towards the boundary of the core. We note these features of the bottom conduction band state and the top valence band  state found in the core region resemble closely the inherent properties of the freestanding [111]-oriented InAs and GaSb nanowires found in the previous studies.\cite{liao2015electronic2,liao2015electronic1} 

It has now become clear that in the shell the electron-like state is more delocalized than the hole-like state, while in the core the opposite is true, i.e., the hole-like state is more delocalized than the electron-like state. Therefore, the electron-hole hybridization is stronger in a GaSb/InAs core-shell nanowire than in the corresponding InAs/GaSb core-shell nanowire.  We emphasize that in other nanowire systems we have seen similar localization characteristics of the electron and hole states.\cite{liao2015electronic2,liao2015electronic1} We therefore speculate that putting the hole state into the core while the electron state in the shell may be a common rule for achieving more robust topologically nontrivial properties in many types of core-shell nanowires.

\section{conclusions}

In this paper, we have presented a detailed study the properties of the energy gaps, especially the topological nature of the fundamental gaps, of the [111]-oriented InAs/GaSb and GaSb/InAs core-shell nanowires. The evolutions of the energy gaps with the structural parameters are shown to be dominantly governed by the quantum size effects. With a fixed core radius, a net, topologically nontrivial fundamental gap exists only in a finite range of intermediate shell thickness, which is true for both InAs/GaSb and GaSb/InAs core-shell nanowires. We have offered detailed maps of the topologically nontrivial fundamental gaps over a wide range of the structural parameters for the two types of core-shell nanowires. Comparatively, we find the electron-hole hybridization is stronger in the GaSb/InAs nanowires due to the more delocalized features of the electron and hole wave functions in the shell and the core, respectively. This explains the larger size of the maximum topologically nontrivial fundamental gap, as well as the persistence of the gap within a wider range of structural parameters in the GaSb/InAs core-shell nanowires. Since the similar localization characteristics of the conduction band electron wave functions and the valence band hole wave functions exist in other types of III-V semiconductor nanowires, it could be used as a general guidance to put the hole-like state into the core while the electron-like state in the shell in constructing a topological core-shell nanowire with a large and robust topologically nontrivial fundamental gap.

\section{Acknowledgments}

This work was supported by the National Basic Research Program of China
(Grants No.~2012CB932703 and No.~2012CB932700) and the National Natural Science Foundation of China (Grants No.~91221202, No.~91421303, and No.~61321001). HQX also acknowledges financial support from the Swedish Research Council (VR).

\bibliographystyle{apsrev}

\end{document}